\documentclass[11pt]{article}
\usepackage{graphicx}

\begin{document}

\newcommand\eq[1]{Eq.~(\ref{eq:#1})}
\thispagestyle{empty}
\begin{titlepage}
\begin{flushright}

\end{flushright}
\vspace{1.0cm}
\begin{center}
{\LARGE \bf The Origin of Fat Tails }\\ 
\bigskip
~\\
\bigskip
\bigskip
{Martin Gremm } \\
~\\
\noindent
{\it\ignorespaces
Pivot Point Advisors, LLC\\
 5959 West Loop South, Suite 333,  Bellaire, TX 77401 \\

{\tt  gremm@pivotpointadvisors.com }

}\bigskip
\end{center}
\vspace{1cm}
\begin{abstract}
We propose a random walk model of asset returns where the parameters depend on market stress. Stress is measured by, e.g., the value of an implied volatility index. We show that model parameters including standard deviations and correlations can be estimated robustly and that all distributions are approximately normal. Fat tails in observed distributions occur because time series sample different stress levels and therefore different normal distributions.  This provides a quantitative description of the observed distribution including the fat tails. We discuss simple applications in risk management and portfolio construction. 
  
\end{abstract}
\vfill
\end{titlepage}

\section{Introduction}

Changes in the market price of securities are thought to be governed by stochastic processes. This idea was first introduced by Bachelier \cite{bachelier} and later rediscovered and refined by many authors (see references in \cite{wilmott} or \cite{merton}). 

The simplest and most popular stochastic model is a (log)normal random walk. In physics it describes Brownian Motion,  the movement of a dust particle as fluid molecules collide with it. In finance it should describe how the price of a security changes as market participants move it up and down by buying and selling. Unfortunately, observed asset returns fail to behave as expected. Instead of following a normal distribution, return  distributions have fat tails. 

This is a problem because the Central Limit Theorem (CLT) states that the distribution of returns should be normal if these returns are the sum of returns on shorter time scales. Financial markets must violate one of the assumptions of the CLT to produce fat tails. Since the CLT makes very few assumptions, there are only a few ways to sidestep it. If financial time series involve more than one distribution (see \cite{engle} for an overview),  certain behavioral patterns \cite{herd}, or autocorrelations they can have fat tails. They can also have fat tails if the returns at short time scales come from Mandelbrot's power law models  \cite{mandelbrot}. 

Many of these models reproduce statistical properties of observed market data including fat tails and volatility clustering, but most do not explain why financial markets should be described by a process that sidesteps the CLT. Volatility clustering and fat tails are generic features of financial markets. They exist throughout recorded history, and they occur in all  asset classes and geographic locations. It is plausible that the explanation for such ubiquitous and durable features should be something equally ubiquitous and durable.

There is strong empirical evidence to suggest that a given level of stress translates into a certain behavior. For example, a \lq{}Flight to Quality\rq{} happens very predictably whenever market participants are stressed. We simply generalize this observation to say that market behavior should generally be a function of market stress. 

Our starting point is a simplification of the standard stochastic volatility model. Instead of allowing the volatility to vary randomly, we assume that it is a function of the stress level of market participants. This is very similar to Brownian Motion in physics, where the standard deviation of the random walk depends on the temperature of the fluid. A dust particle in a warmer fluid moves  more than one in a cooler fluid. 

We propose that market stress serves the same function as temperature in Brownian Motion.  Stressed market participants should make larger and more frequent trades than calm ones. This increase in trading activity should result in higher volatility of asset prices. As in Brownian Motion, we should find a log-normal random walk at any given level of market stress. 

A time series of returns samples various stress levels and therefore different normal distributions. Such models are known to generate time series with fat tails. To specify the model we only need to estimate the functional dependence of the mean and standard deviation on stress levels. As we will show, this is easy to do because stress levels are directly observable in the form of, e.g., implied volatility indices.

In the next section we define our model and discuss its general properties. In Sect.~\ref{singleassetfit} we use market data for stock and bond indices to verify the predictions of our model and to estimate its parameters. A brief detour in Sect.~\ref{musection} confirms the plausible expectation that  changes in stress levels determine market direction. 

In our framework it is easy to estimate the probability of extreme events. We discuss this in Sect.~\ref{risk}. Sect.~\ref{multiassetportfolios} expands the discussion to an example of a two asset portfolio. Sect.~\ref{noniid} briefly discusses concepts, such as Sharpe Ratios and Modern Portfolio Theory, which need to be rethought in our framework. Sect.~\ref{conclusions} summarizes our results. 

\section{The Model}
\label{model}

The standard log-normal random walk  is given by the  stochastic differential equation

\begin{equation}
\label{simplemodel}
\frac{dS}{S} = \mu dt+\sigma dX.
\end{equation}
Here $S$ is the asset price, $\mu$ is the mean, $\sigma$ is the standard deviation, and $dX$ is a normally distributed stochastic process with zero mean and unit standard deviation. $\mu$ and $\sigma$ are assumed to be constants. 

We propose to replace these two constants with functions that depend on $\kappa(t)$,  the stress level at $t$. The model then reads

\begin{equation}
\frac{dS}{S} = \mu(\kappa(t)) dt+\sigma(\kappa(t)) dX.
\end{equation}
 
Implied volatilities are a good indicator of market stress because they effectively measure the price of insurance against adverse market moves.  For example, we can use VXO, an index that measures the implied volatility of certain OEX options, as a proxy for stress in the equity market.  $\kappa(t)$ is simply the index value at time $t$.  More generally, if an implied volatility index exists for an asset class, it provides a direct way of measuring stress in that market\footnote{There are other proxies for market stress besides implied volatilities. We found that using credit spreads for bonds, and trailing observed volatilities for stock and bonds produces qualitatively similar results.}.

Assuming that $\mu$ and $\sigma$ depend on $\kappa$ is a very minimal extension of Bachelier's original model \cite{bachelier} because it does not add any new degrees of freedom. $\kappa(t)$ is a measurable function that labels historical observations with a stress level.

It is helpful to list a few general predictions of our framework before we delve into detailed data analysis in the next section. 

\begin{itemize}

\item
As we discussed in the introduction, the rationale behind introducing a model with stress-dependent parameters is that stressed market participants are expected to make larger and more frequent trades than unstressed ones. We should see trading volumes increase with rising stress levels.

\item

Our model should work for all liquid asset classes in the sense that investors should show qualitatively similar responses to  stress. For example, a stressed stock investor should  behave qualitatively the same way as a stressed bond investor, but these two investors will generally not be stressed at the same time. 

\item
The  CLT demands that returns at a given value of $\kappa$ should follow a normal distribution. Our model leaves no room for any of the standard ways to sidestep it.

\item
Estimates of standard deviations and correlations from market data are very unstable. In our model this is expected behavior because different data samples contain different combinations of stress levels. We should see that estimates from  observations at a given level of $\kappa$ are stable and vary smoothly with  $\kappa$. 

\end{itemize}

In the next section we estimate the parameters of our model for two asset classes and verify these general expectations.

\section{Single Asset Data Survey}
\label{singleassetfit}

Our data set contains daily continuously compounded returns for the  S\&P 500 Index (SPX) and the Ryan Labs 10 Year Bond Index (RT10) from 4/28/1988 to 8/13/2012. Each day is labeled with the closing values of the VXO\footnote{The VXO is similar to the VIX, but has a longer track record.} and MOVE indices, which measure the implied volatility of certain OEX and treasury bond options respectively. These two indices are our $\kappa(t)$ for the stock and bond market. Our dataset also contains the detrended trading volume for the S\&P 500.

\begin{figure}[h]
\center
\includegraphics[width=120mm]{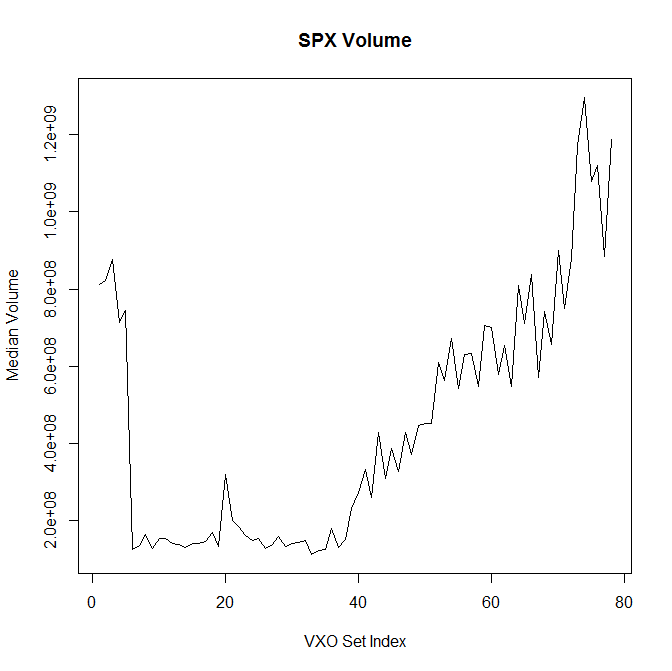}
\caption{S\&P 500 Trade Volume vs.  Stress}
\label{volume}
\end{figure}

For the analysis in this section we  group observations into non-overlapping sets of 75 observations. Unless noted otherwise, charts plot sample estimates from these sets against the sequential index of the set.

Before we explore our model we need to verify our central assumption that trading volumes increase with stress levels. Fig.~\ref{volume} shows the median daily volume calculated from observations ordered by ascending $\kappa$.

Trading volumes are fairly constant across most of the lower stress part of the plot except at the very lowest stress levels where trading volumes rise sharply. Most of those data points come from the late-stage bull market in 2006 and early 2007, reflecting  the tendency of investors invest heavily into the end of a bull market. It is plausible that factors other than stress determine trading volumes in calm markets. The right half of the plot shows the expected behavior of rising trading volume as stress increases.

\begin{figure}[h]
\center
\includegraphics[width=120mm]{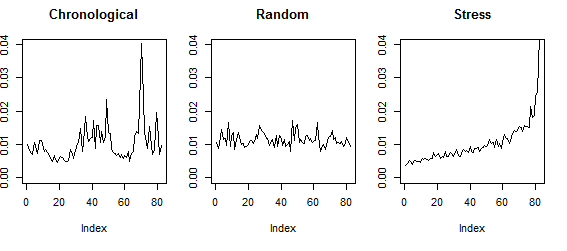}
\caption{S\&P 500 Sample Volatility}
\label{vxovol}
\end{figure}

Let us now turn to analyzing other market data as a function of stress levels. The left chart in Fig.~\ref{vxovol} shows the standard deviation of the S\&P 500 with the returns ordered chronologically. Sample estimates are clearly not stable. There are extended periods of higher and lower sample volatility. This is commonly referred to as volatility clustering.

The middle chart shows estimates after randomizing the order of returns. If the returns were an independent and identically distributed (IID) sequence, this reordering should have no effects. The difference between the left and the middle chart illustrates that the return time series is not IID.

The chart on the right shows standard deviations calculated from returns sorted by increasing $\kappa$. If the returns come from a stochastic process that has no $\kappa$-dependence, sorting  the returns by $\kappa$ should have the same effect as ordering them randomly. We see that sample estimates rise smoothly with $\kappa$ and that the variability of sample estimates is sharply lower than for the two other charts. These observations strongly suggest that the parameters of this stochastic process  depend on $\kappa$.

The chart on the left shows more structure than the randomized one because market stress often runs high for extended period during  crises before falling back down to non-crisis levels. This means that chronological returns represent a partial stress ordering. It also explains volatility clustering observed in financial time series.

\begin{figure}[h]
\center
\includegraphics[width=120mm]{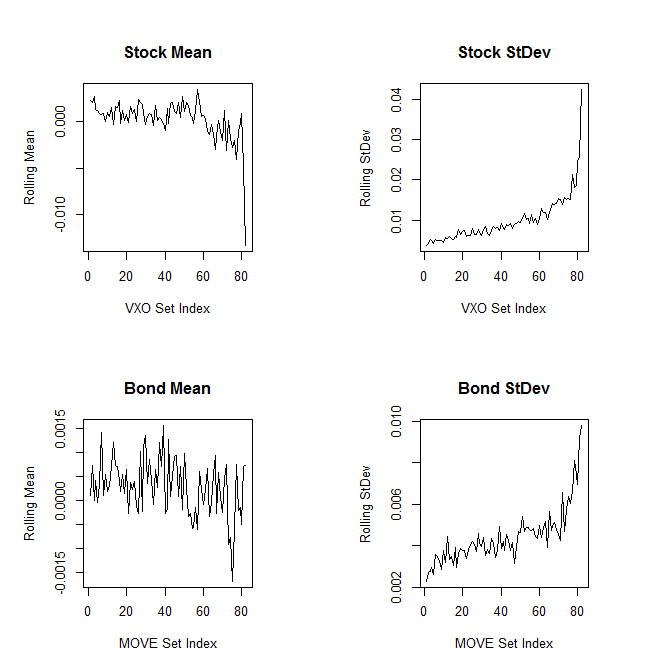}
\caption{Stress-Ordered Sample Estimates}
\label{stressordered}
\end{figure}

Fig.~\ref{stressordered} shows the sample estimates of $\mu$ and $\sigma$ as functions of $\kappa$ for stocks and bonds. In both cases we find that $\sigma$ increases smoothly with $\kappa$ while $\mu$ shows little sensitivity except for the days when the stock market is most stressed. We discuss $\mu$ in more detail in Section~\ref{musection}. 

This confirms two of our expectations listed in the previous section. Different asset classes show qualitatively the same behavior and stress ordering returns results in stable sample estimates, at least for the standard deviation. Let us now turn to the third prediction that the returns for a given level of $\kappa$ are approximately normally distributed.

In practice we need to sample a range around the desired value of $\kappa$ in order to have enough observations for a reasonable estimate. This means that we are sampling distributions with similar but not identical standard deviations. This will modify the tails of the observed distribution, but the effect should be small since most of our 75-observation sets cover only a small range of $\kappa$.

\begin{figure}[h]
\center
\includegraphics[width=120mm]{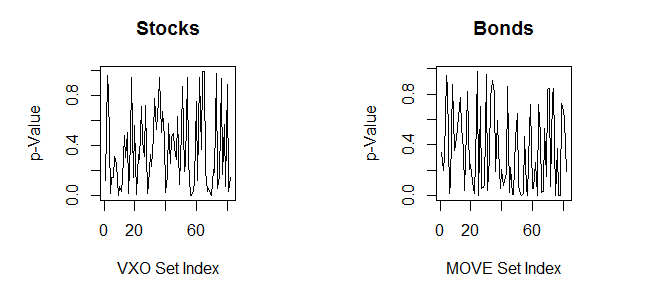}
\caption{p-Values for the Shapiro Wilk test for normality}
\label{pvalue}
\end{figure}
  
Fig.~\ref{pvalue} shows the p-Values of the Shapiro Wilk test for normality, calculated for each non-overlapping set of 75 observations. A simple way to test our expectation that the returns in each set should be normally distributed is to examine the distribution of p-Values.  For example, we can count the number of sets with a p-Value below 0.05. The returns in these sets have a greater than 95\% probability of coming from something other than a normal distribution. 

If all of our samples actually do come from normal distributions, we would expect to see false positives (i.e., returns that look non-normal with a 95\% confidence level)  in 5\% of the samples. The actual numbers are higher, 17\% for stocks and 22\% of bonds.  Some increase was expected because we are sampling a range of $\kappa$. In the next section we briefly discuss another, possibly larger contribution to this residual non-normality. 

Returns are substantially more normal after partitioning them by $\kappa$ compared to ordering them randomly. 83\% of the random  sets look non-normal at a 95\% confidence level for stocks. For bonds the number is 44\%. For completeness we also quote  chronologically ordered results. Since this is a partial stress ordering, they fall between the stress-ordered and the randomly ordered results. We find that 28\% of the stock sets and 26\% of the bond sets look non-normal using chronological returns. 

\begin{figure}[h]
\center
\includegraphics[width=120mm]{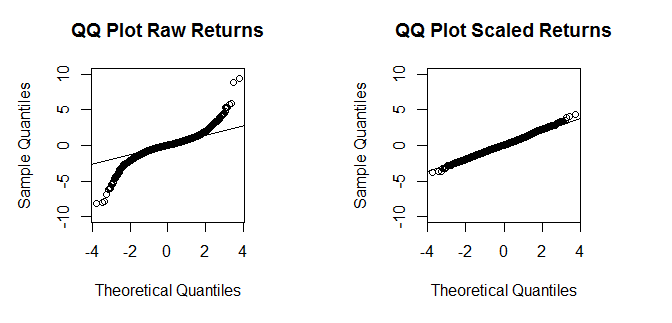}
\caption{Chronological and Rescaled SPX Returns}
\label{qq}
\end{figure}

In order to visualize the residual non-normality we can compare the distribution of the chronological returns to the distribution of chronological returns after dividing each return by $\sigma(\kappa)$. If our model explained all of the non-normality in our data, this would transform the historical time series into a normal distribution with unit standard deviation. 

Fig.~\ref{qq} shows that our model accounts for most of the non-normality of observed returns. The residual non-normality indicates that other factors also play a role, but that changing stress levels appear to be the dominant factor. It is plausible that it is not the only factor. $\sigma$ should also depend on the actions of the Federal Reserve, the depth of the order book, options expiry, and countless other factors. 

The results in this section confirm the three general predictions from the previous section: qualitatively similar behavior across asset classes, stable sample estimates, and approximately normal return distributions at constant stress.

\section{What Determines the Mean?}
\label{musection}

In the previous section we saw that standard deviations depend strongly on market stress, but that the means do not. We also found evidence that the parameters of our stochastic model depend on additional factors. 

\begin{figure}[h]
\center
\includegraphics[width=90mm]{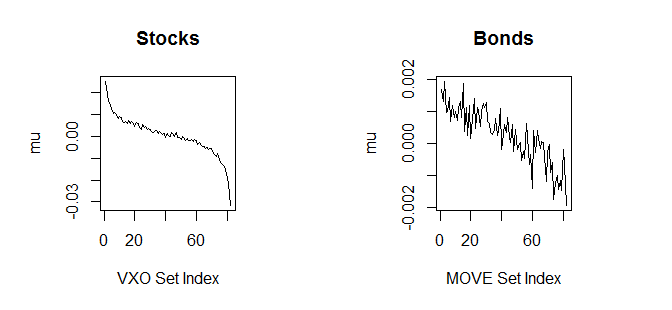}
\caption{$\mu$ as a function of $r_\kappa$}
\label{muplots}
\end{figure}

Let us introduce $r_\kappa$, the one-day percentage change of $\kappa$. Applying the same strategy as before, we sort the returns by $r_\kappa$ and calculate windowed estimates of $\mu(r_\kappa)$. Fig.~\ref{muplots} shows the results. It is clear that the change in stress levels drives returns. If stress increases, returns tend to be negative, if it decreases they tend to be positive. 

This is, of course, no surprise. It is natural for an investor who gets more concerned about an asset to sell it. Our framework simply allows us to quantify the effect.

The $r_\kappa$ dependence of $\mu$ is an example of a relevant, but sub-leading factor that we ignore in this paper for simplicity. 

\section{Single Asset Risk Estimates}
\label{risk}

Risk managers often calculate a Value at Risk (VaR) estimate for their portfolio. VaR measures the probability of losing more than a certain amount over a specified time period. The simplest form of VaR inserts the mean and standard deviation from a historical time series of asset returns into the Cumulative Distribution Function (CDF) for the normal distribution to calculate the probability of a return of less than  $x_0$,

\begin{equation}
CDF(x_0)=   \frac{1}{\sigma \sqrt{2\pi}} \int_{-\infty}^{x_0} dx e^{-\frac{1}{2}\left(\frac{x-\mu}{\sigma}\right)^2}.
 \end{equation}

This approach has many shortcomings. Sample estimates of $\mu$ and $\sigma$ tend to be unstable (see Fig.~\ref{vxovol}), which limits the predictive power of VaR, and VaR ignores the fat tails of observed return distributions because the probability distribution is assumed to be normal. More sophisticated ways to estimate VaR exist, but most of them suffer to some degree from the same problems as the basic approach. 

Since VaR calculations often do not account for the fat tails of return distributions, risk managers tend to supplement them with scenario analyses to see how the portfolio would have done during historical or hypothetical crises. These estimates are useful examples of how a portfolio might behave, but this approach does not provide a framework for estimating the probability of stress events, especially ones that have not yet been observed.

Our model avoids many of these problems. Sample estimates at a given stress level are very stable over time, so we can use them reliably to predict future behavior. We showed in Sect.~\ref{singleassetfit} that the distributions in our model are (close to) normal. The fat tails that bedevil traditional risk estimates  emerge only because chronological returns sample different stress levels. Since this process is transparent in our model, we can calculate the probability and magnitude of \lq{}tail events\rq{} by making VaR-equivalent risk estimates. 

In our model the CDF takes the following form:

\begin{equation}
\label{cdf}
CDF(x_0) = \int_0^\infty d\kappa P(\kappa)  \frac{1}{\sigma(\kappa) \sqrt{2\pi}} \int_{-\infty}^{x_0} dx e^{-\frac{1}{2}\left(\frac{x-\mu(\kappa)}{\sigma(\kappa)}\right)^2},
\end{equation}
where $P(\kappa)$ is the probability distribution of the stress levels. The integral over $\kappa$ adds up the probability of a return below $x_0$ for every level of $\kappa$. 

The CDF provides an easy way to test our model out of sample. We already know that estimates from adjacent 75-observation sets yield very similar results. It should also be true that parameters estimated from a set of market data should accurately predict the distributional properties of another dataset.

To test this we randomly subdivide our dataset into two sets of roughly equal size. We then use one of the subsets to estimate $P(\kappa)$, $\mu(\kappa)$ and $\sigma(\kappa)$  for use in the CDF. The distribution of the other dataset should conform to this CDF.

A simple frequency analysis provides a rough estimate of $P(\kappa)$. We assume that the training dataset is large enough to yield a good sampling of stress levels. For a time series that contains some calm periods and a few good crises this should be the case.

\begin{table}[h]
\center
    \begin{tabular}{|l|r|r|r|}
        \hline
        $\kappa$/VXO & $P(\kappa)$ &  $\mu$ &  $\sigma$ \\ \hline
        00-10  & 0.8\%               & 0.00288    & 0.00310        \\ 
        10-20  & 52.3\%               & 0.00097    & 0.00679        \\ 
        20-30  & 34.9\%               & 0.00052       & 0.01163          \\ 
        30-40  & 8.5\%                & 0.00010      & 0.01761          \\ 
        40-50  & 2.5\%                & -0.00495      & 0.02634          \\ 
        50-60  & 0.2\%                & -0.03426      & 0.04302           \\
        60-70  & 0.4\%                & 0.00598      & 0.05707           \\
        70-max  & 0.3\%                & -0.03952      & 0.04146           \\
        \hline
    \end{tabular}
    \caption{S\&P 500 sample estimates.}
\label{pkappatable}
\end{table}

\begin{figure}[h]
\center
\includegraphics[width=120mm]{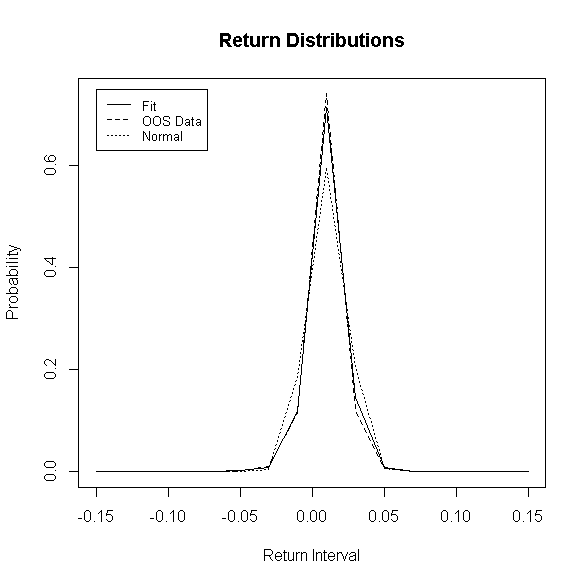}
\caption{S\&P 500 Return Distributions. Fit is the prediction, OOS is the out of sample observation, Normal is the prediction if we fit a normal distribution to the sample data.}
\label{distributions}
\end{figure}

Using these results and Eq.~\ref{cdf} we calculate the percentage of observations that should fall into various return brackets. Fig.~\ref{distributions} shows that the out-of-sample data matches the expected distribution function very closely. The plot also includes the expected return distribution if we fit a normal with fixed $\mu$ and $\sigma$ to the data. The distribution using our model has significantly fatter tails than a normal distribution.

Armed with the results in Table \ref{pkappatable} we can calculate several risk estimates for the S\&P 500. The first is the probability of experiencing a one-day decline of at least 9\%, which is a typical magnitude for market crashes. We use  Eq.~\ref{cdf} to calculate the probability of $8\times 10^{-4}$  (once in about five years) of a decline worse than 9\%. The traditional VaR approach gives $2\times 10^{-14}$. Our dataset contains about 24 years of data and three days with returns below $-9\%$, roughly in line with our expectations.

The second is the probability of experiencing another event like the crash of \lq{}87 when the S\&P 500 declined 22.9\%.  The probabilities for a decline worse than 22.9\% are $9\times 10^{-8}$ (roughly once in 44,000 years) and $4\times 10^{-82}$ respectively. The crash of \lq{}87 is an outlandish event in both approaches, but in our approach the probability of such an event is 74 orders of magnitude greater than in the traditional VaR calculation, which changes the fact that we observed such an event from virtually impossible to merely extraordinarily unlikely.

We can also make a pessimistic risk estimate by assuming that we will encounter a stressed environment. For example, we can estimate risks assuming that the VXO index is above 70. With these assumptions, the probability of encountering a decline of more than 9\% is about 11\%. The likelihood of another crash of \lq{}87 is $2.4\times10^{-6}$.

The bond data exhibits the same qualitative behavior. The tails are much fatter than those of a normal distribution. For example, in our framework the probability of a return between -3\% and -2.5\% is $1.3\times 10^{-4}$ while a normal fit predicts $3.8\times 10^{-8}$. Here too the out of sample data matches our predicted distribution. 

$\kappa$, has strong autocorrelations because stress levels normally do not change dramatically from one day to the next. Crises tend to last for weeks or months and calm periods also often extend for months or even years. We can make a stress estimate for tomorrow using a predictive model for $\kappa$ or assuming that $\kappa$ will remain approximately the same. Most of the time, this estimate will be closer to observed reality than the agnostic estimate in Eq.~(\ref{cdf}). It will also be dramatically wrong when stress levels do change significantly.

\begin{figure}[h]
\center
\label{forecastrescaling}
\includegraphics[width=120mm]{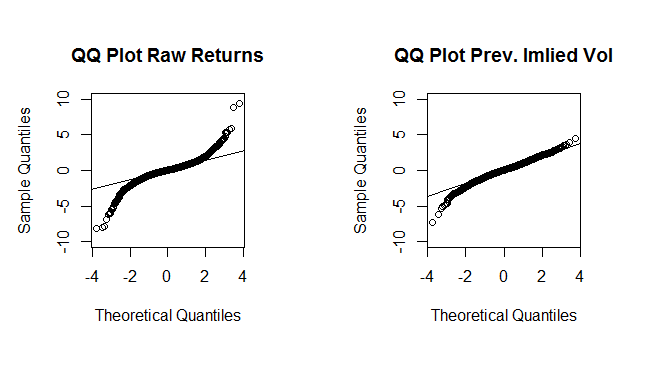}
\caption{Chronological and Rescaled SPX Returns}
\end{figure}

To illustrate this, let us forecast the next day's $\kappa$ by assuming it is the same as today. Armed with this forecast we construct another rescaled time series as we did for Fig.~\ref{qq}, but this time using our forecast for $\kappa$ to find the next day's projected $\sigma(\kappa)$. Fig.~\ref{forecastrescaling} shows the result. Since our forecast is imperfect, this rescaling is less effective than one in  Fig.~\ref{qq}. However, stress levels stay approximately constant for extended periods, so the rescaled time series is more normal than the unrescaled one.

These examples illustrate how risk can be quantified in our model. The same method yields probabilities for any return from the smallest to the most extreme. So-called tail events become ordinary returns that do not require special treatment.

\section{Multi Assets Portfolios}
\label{multiassetportfolios}

\begin{table}[h]
\center
    \begin{tabular}{c c}
	\includegraphics[width=2.25in]{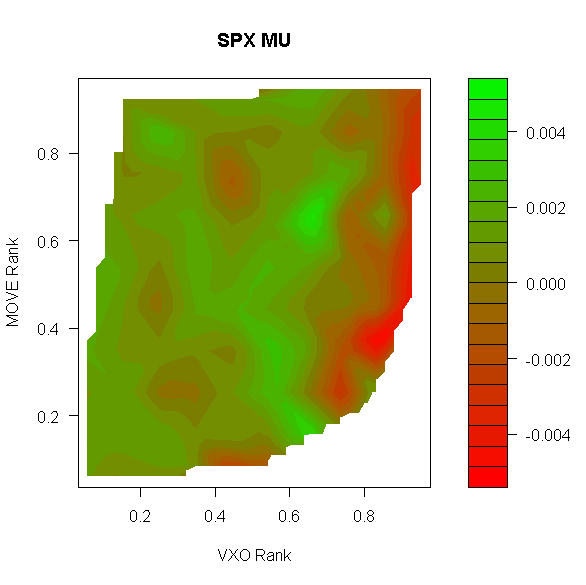}	&	\includegraphics[width=2.25in]{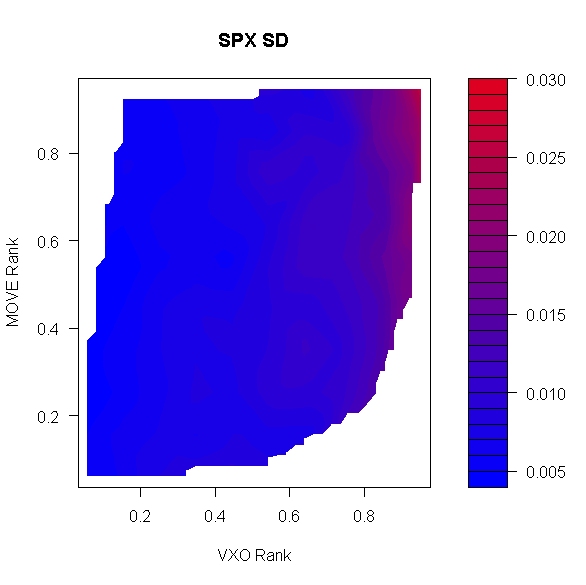}	\\ 
	\includegraphics[width=2.25in]{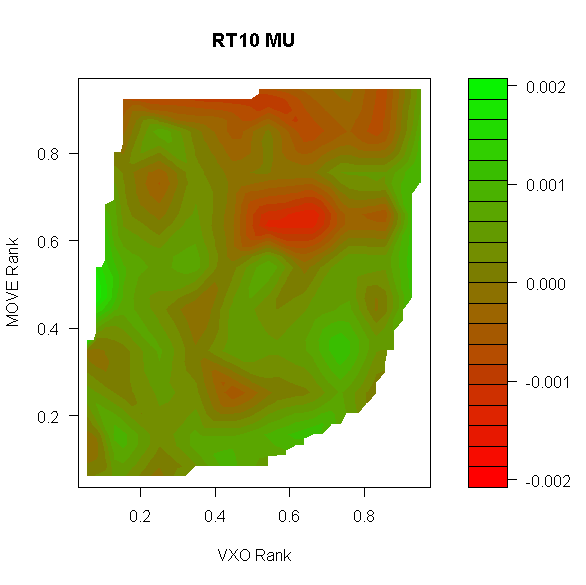}	&	\includegraphics[width=2.25in]{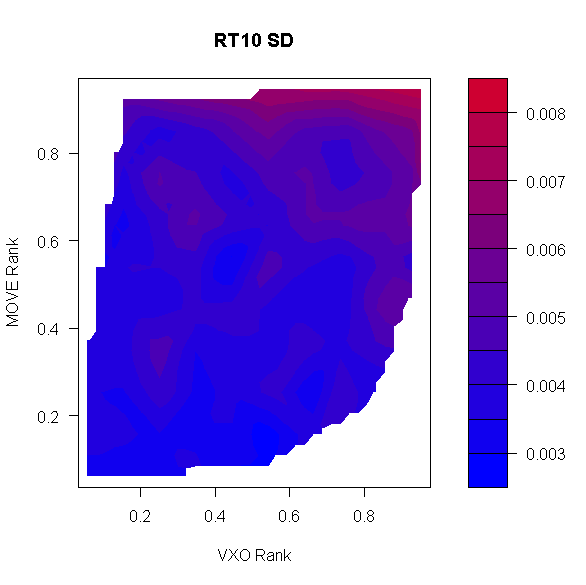}	\\ 
      \end{tabular}
      \caption{Two dimensional sample estimates of $\mu_{s/b}$ and $\sigma_{s/b}$}
\end{table}

In this section we consider portfolios that contain stocks and bonds as represented by the indices in our dataset. This requires a small generalization of our original model. Since the stock market is not necessarily stressed at the same time as the bond market, we need to partition our data by the stress levels of both the stock and bond market. This changes our model to

\begin{equation}
\frac{dS_{s/b}}{S_{s/b}} = \mu_{s/b}(\kappa_s(t),\kappa_b(t)) dt + \sigma_{s/b}(\kappa_s(t),\kappa_b(t)) dX,
\end{equation}
where $\kappa_{s/b}$ is the stress level of the stock and bond market respectively.

We divide our dataset into a 10 x 10 grid by $\kappa_{s/b}$ to estimate $\mu$, $\sigma$, and the correlation coefficient for the two assets, $\rho$, as a function of the stress levels. The observations in each square fall into a certain decile of stock and bond stress. Not all squares contain observations. For example, there are no observations where the VXO value falls into the top decile and the MOVE value falls into the bottom decile. 

The results are  what we should expect based on the single asset analysis in the previous section and empirical observations. For stocks and bonds, the standard deviations increase with the corresponding stress. An increase in stock stress gives a small boost to bond volatility. Expected returns show little sensitivity to stress levels except at the extremes of the corresponding stress measure where they turn negative. The sensitivity of $\mu$ to stress levels is stronger for stocks than for bonds. 

\begin{figure}[h]
\center
\includegraphics[width=100mm]{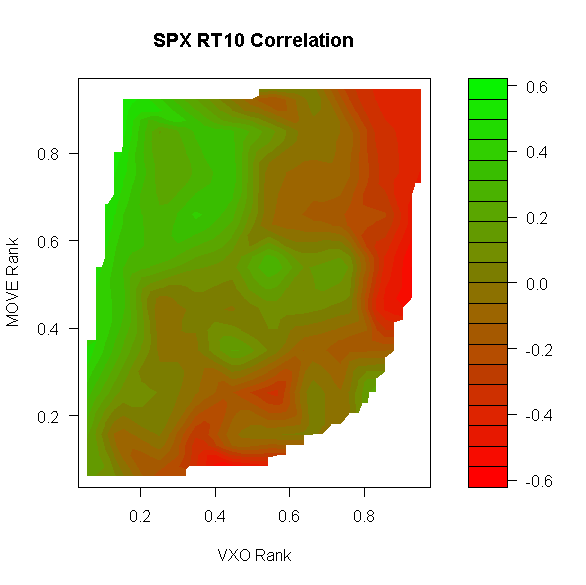}
\caption{Stock/Bond Correlation Coefficient}
\label{cor}
\end{figure}

The correlation between stock and bond returns is  strongly positive in fairly large parts of the  parameter space. Fig.~\ref{cor} shows that the correlation is negative when the stock market is stressed, but that it turns positive when stock stress is low, even when bond stress is high. 

We can use these results to analyze the behavior of portfolios containing stocks and bonds be generalizing Eq.~\ref{cdf} to the two-dimensional case with a joint probability distribution of stock and bond stress, $P(\kappa_s,\kappa_b)$. The findings are qualitatively similar to the one-dimensional case, so we will not discuss them here. As in the one-dimensional case our framework provides the tools to estimate tail risk for multi-asset portfolios rigorously and robustly.

The main reason for investing into uncorrelated or negatively correlated assets is the diversification benefit. Conventional wisdom posits that a portfolio containing stocks and bonds is better than one that contains only one or the other, because they tend to move in opposite directions.

The analysis in this section shows that this is only true under certain market conditions because the correlation between stocks and bonds is not always negative. A stock investor gains diversification by adding bonds to the portfolio because the correlation between the two assets is negative when the stock market is stressed. This is desirable because bonds have a positive return in times of stock stress and they reduce the volatility of the whole portfolio. 

A bond investor obtains a diversification  benefit only when both the stock and the bond markets are stressed. Even then it is not clear that this diversification is desirable, because the stock market tends to produce negative returns when it is stressed.

This example shows that building diversified portfolios becomes more complicated, but probably also more successful, if we take into account that markets change with stress levels. 

\section{Inconstant Markets}
\label{noniid}

Many concepts in finance assume that asset returns are IID. Our analysis indicates that this is not the case and that the properties of markets change quite dramatically depending on stress levels. In this section we briefly discuss a few examples of concepts that need to be rethought in our framework. 

\begin{figure}[h]
\center
\includegraphics[width=90mm]{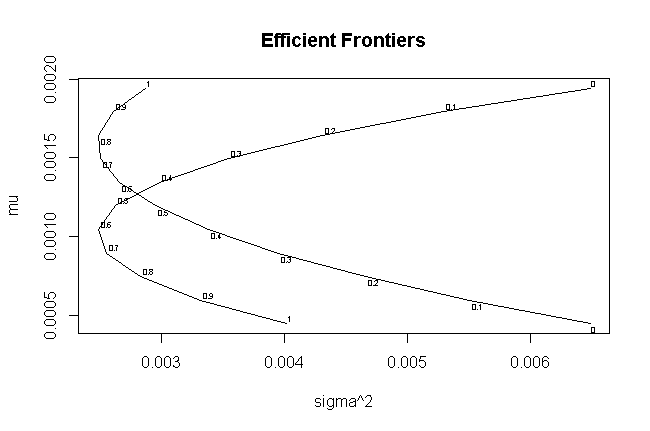}
\caption{An Efficient and an Inefficient Frontier}
\label{ef}
\end{figure}

The results of the previous section have implications for Modern Portfolio Theory (MPT). MPT uses estimates of $\mu$ and $\sigma$ for several assets to identify optimal portfolios in terms of risk adjusted returns. These portfolios are said to reside on efficient frontiers, which are normally plotted on a coordinate system with $\sigma^2$ on the horizontal axis and $\mu$ on the vertical axis. Each asset is represented by a point in this plane. The two most attractive ones define a parabola. The arm with the higher return is said to be the efficient frontier.  Fig.~\ref{ef} shows an example.

In our framework we find a different efficient frontier for every value of $\kappa_{s/b}$. Fig.~\ref{ef} shows two examples for a portfolio of stocks and bonds.  The bottom parabola describes a time of low stock and bond stress. Both assets have positive $\mu$ and the correlation is positive. The curves are labeled with the percentage of bonds in the portfolio. An efficient portfolio for these market conditions has a large exposure to stock.

The second parabola uses $\mu$ and $\sigma$ from a high stress point. The correlation is negative and so is $\mu_s$. Most of the weights that were on the attractive side of the unstressed parabola have migrated to the unattractive side of the stressed one. Only portfolios that are almost exclusively bonds are on the efficient frontier in stressed markets. It is clear that efficiency depends on market conditions.

Another  example is the Sharpe Ratio, a measure of risk adjusted return of an asset given by the expected (or historical) return less the risk free rate divided by the expected (or historical) risk. Sharpe Ratios almost always use the standard deviation to measure risk. This gives informative results if the returns are normally distributed. If the distribution has fat tails, the standard deviation underestimates the real risk.

In our framework the Sharpe Ratio is only well-defined for a given value of $\kappa$. For practical purposes we need to measure risk adjusted returns over extended periods of time. Since such time series sample different stress levels, the Sharpe Ratio cannot be calculated.

We propose a measure of risk adjusted return that accommodates fat tails and explicitly accounts for the variability of stress levels: The probability of losing more than the expected accumulated returns over a fixed time period. The expected $N$-day return is

\begin{equation}
r_N=N \int_0^\infty d\kappa P(\kappa)\mu(\kappa).
\end{equation}

Using Eq.~\ref{cdf}, we can calculate the probability of a one-day return that wipes out more than $N$ days of expected return, $P_{N}=CDF(-r_N)$. This concept makes no implicit distributional assumption and it remains well defined when $\mu$ and $\sigma$ depend on $\kappa$. 

We find $P_{100}=0.0059$ for stocks and $P_{100}=0.0002$ for bonds. The rather large difference reflects the much fatter tails of the stock distribution. For comparison Sharpe Ratios calculated from the whole data set are $0.024$ for stocks and $0.060$ for bonds. These values are much closer because the Sharpe Ratio does not account for fat tails.

\begin{figure}[h]
\center
\includegraphics[width=90mm]{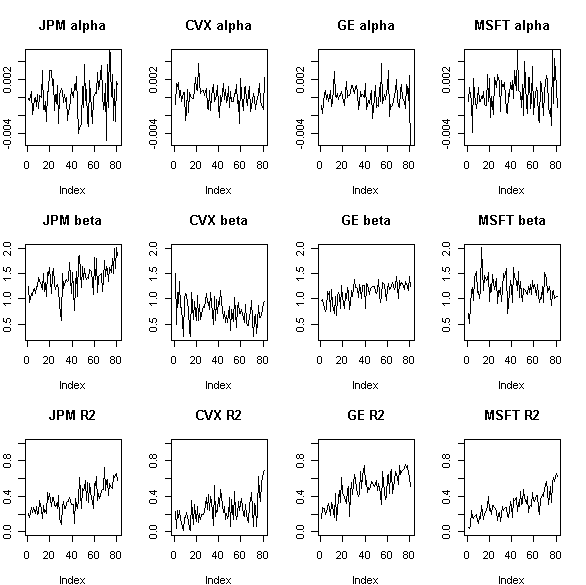}
\caption{Regression coefficients and $R^2$ for stocks vs. the S\&P 500}
\label{r2}
\end{figure}

The Capital Asset Pricing Model (CAPM), like the two previous examples, implicitly assumes that there is a constant market return and constant regression coefficients of stock prices against a market benchmark. Once again, in our framework these quantities become functions of $\kappa$. Fig.~\ref{r2} shows some examples. 

The general behavior is what we should expect based on the single asset analysis in Section \ref{singleassetfit}. The intercept ($\alpha$) is rather insensitive to stress levels, while $\beta$  changes as stress levels rise. Interestingly, $\beta$ for JP Morgan and General Electric rise with increasing stress while they decrease slightly for Chevron and Microsoft. This suggests that the first two are unsuitable stocks to hold in a crisis because they amplify market volatility more and more as stress levels rise. 

The bottom row in Fig.~\ref{r2} shows that the $R^2$ increases as stress levels rise. This provides quantitative confirmation of the well-known fact that in stressed markets all stocks move together so that the overall market direction is a very good predictor of individual stocks\rq{} moves. In unstressed markets, stock moves are driven by stock specifics. Overall market direction is only one of many factors affecting returns.

Anything that depends on means, standard deviations, correlations, regression coefficients, or fat tails can be reexamined in this framework. This section provides a few simple examples to give a flavor of how familiar concepts change in our model.

\section{Conclusions}
\label{conclusions}

This paper offers evidence that the stock and bond markets are described by simple log-normal stochastic processes with stress-dependent parameters. This model is a formal description of the empirical fact that markets behave differently when investors are fearful than when they are calm. 

We show that grouping market data by stress eliminates or explains many of the oddities of financial time series including the otherwise implausibly fat tails of return distributions, unstable sample estimates correlations and standard deviations, and volatility clustering. A few simple examples to illustrate how our framework can be used for risk management and to explore financial data. 

The purpose of this paper is to propose the idea and to supply some supporting  evidence. The next step it to test these ideas more rigorously. If the model survives, many familiar concepts in finance from risk management the efficient market hypothesis will need to be rethought.

\vskip0.1in

\end{document}